This is an author-generated version.

The final publication is available at link.springer.org

DOI: 10.1007/11754305_36
Link: http://link.springer.com/chapter/10.1007%2F11754305_36

Bibliographic information:

Alexis Ocampo, Jürgen Münch. Process Evolution Supported by Rationale: An Empirical Investigation of Process Changes. In Software Process Change, volume 3966 of Lecture Notes in Computer Science, pages 334-341, Springer Berlin Heidelberg, 2006.

# Process Evolution Supported by Rationale:
# An Empirical Investigation of Process Changes

Alexis Ocampo, Jürgen Münch

Fraunhofer Institute for Experimental Software Engineering
Fraunhofer-Platz 1, 67663, Kaiserslautern, Germany
{ocampo, münch}@iese.fraunhofer.de

**Abstract.** Evolving a software process model without a retrospective and, in consequence, without an understanding of the process evolution, can lead to severe problems for the software development organization, e.g., inefficient performance as a consequence of the arbitrary introduction of changes or difficulty in demonstrating compliance to a given standard. Capturing information on the rationale behind changes can provide a means for better understanding process evolution. This article presents the results of an exploratory study with the goal of understanding the nature of process changes in a given context. It presents the most important issues that motivated process engineers changing important aerospace software process standards during an industrial project. The study is part of research work intended to incrementally define a systematic mechanism for process evolution supported by rationale information.

## Introduction

Software process models are used as a means for supporting software engineers in systematically performing the engineering processes needed to develop software products. As these processes are performed, suggestions for adjustment or refinement can arise, which in turn demand evolving the models. Usually, certain events such as the introduction of a new software development technology in a development team (e.g., new testing support tools and techniques), a new/updated process engineering technology (e.g., a new process modeling technique), new/updated standards/guidelines for software development or process engineering, new/updated regulatory constraints, or new/updated best practices emerging from community experience generate issues that must be resolved by performing changes to the software process models.

In many cases, precipitous and arbitrary decisions are taken, and process models are evolved without storing or keeping track of the rationale behind such changes. One of the reasons is that this is an expensive activity that demands a dedicated role in the organization [1], especially because identifying the rationale of a change, or driving evolution activities in terms of rationale, is not an easy task. A mechanism (concept and tool) that can be used for collecting information about process changes and that could help in evolving the process in a systematic way is needed.

We believe that the first step towards such a systematic mechanism is to understand the nature of process changes. We assume that by having a predefined classification of the most common reasons for process changes, the task of collecting the information related to such a rationale can be simplified and become more suitable for use in real process evolution projects. Additionally, this can be seen as an initial step for building a mechanism that supports systematic process evolution. Once this is understood, a structured conceptual model of rationale can be produced and tested in process evolution projects.

This article presents the results of an initial attempt to achieve such a predefined classification as follows: Section 2.1 briefly presents the basic concepts that we use for understanding the nature of changes to the software process and software process evolution. Section 2.2 provides short descriptions of related work where concepts for understanding process changes have been developed; Section 3 presents the context of the study performed for understanding the nature of changes to a process standard. Section 4 presents the issues derived from a repository of changes performed to a process standard, and an interpretation of the frequency with which such issues appeared during the project. Section 5 presents a discussion of the most relevant findings of the study together with research questions to be addressed in the future.

## Background

### Process Evolution Supported by Rationale

We believe that software process evolution should describe the relationships between an existing process model and its pre-existing version(s). Such relationships denote differences between versions due to distinguishable modifications.

One can distinguish the meaning of such modifications if one can understand the rationale behind them. Rationale is defined as the justification of decisions [1]. Historically, much research about rationale has focused on software/product design [9], [10], [11], and [12]. Rationale models represent the reasoning that leads to the system, including its functionality and its implementation [3]. In general, the capture, organization, and analysis of change rationale appears to be a research topic extensively addressed by software product designers but unknown to, or considered unimportant by, software process engineers. This conflicts with the obvious requirement that process engineers need to know the process evolution history in order to be able to effectively and efficiently tailor processes or update them. For example, tailoring a process model without considering what is or is not suitable for a given project can lead to undesired results. This was observed in the study presented in this paper. Process engineers found through interviews that a tailored process forced process practitioners to take part in system design activities that they felt they did not belong to, especially because this was not part of their work scope. Practitioners assured them that such activities were part of the tailored process although they did not know why, since a previous version of the tailored process did not have them. As a consequence, practitioners and process engineers were all confused and without information that could lead to a suitable solution. Tailoring can be successfully accomplished if a process engineer knows the issues, alternatives, arguments, and criteria that justify the definition of a process model. Equally, updating a process model without having knowledge of its history can lead to process models that do not reflect actual practices. Some other benefits of using the rationale as driver for software process model evolution are: supports reworking of software process standards; supports understanding the impact of changes due to specific issues; encourages making rational decisions instead of emotional ones; supports the analysis and identification of non-systematic and rushed decisions.

**Related Work**

There are not many studies that report on a classification or taxonomy of reasons for changing a process model. Nguyen and Conradi [2] present a framework for categorizing process evolution based on six dimensions (origin, cause, type, how, when, and by whom). A change categorized by this framework is called a change pattern. The change pattern, project characteristics, and product quality attributes are stored together so that they can be used for future projects. Data on the evolution of a software development project were collected in a case study performed in the software development department of a banking institution. With regard to the "where", i.e., the sources of process changes, 40% of the recorded changes were due to customer requests, and 60% were due to changes from senior or middle management. The most common observed reasons (why) were the following: a) misunderstanding originating from the customer; b) resources and competence was not always available; c) a new approach for solving the problem was adopted.

Madhavji [5] presents the Prism model of changes, which is an abstract description of a software environment specialized in the treatment of changes in a software development project. The Prism model serves as a classification scheme for structuring the decisions that change an item and as an information base suitable for analyzing the history changes that can help to make future decisions. Unfortunately, Madhavji [5] does not provide a deeper insight or data that show a classification of reasons.

Bandinelli et al. [4] identify three significant categories of changes caused by a variety of reasons and needs. They are: 1) incremental definition: Processes cannot be completely defined at the beginning of a project; therefore, changing them continuously can be viewed as a type of change that adds new parts to the process model; 2) environmental/organizational: Changes of this type are caused because, e.g., the company has acquired new tools to support the software development staff; 3) customization: Changes of this type allow process agents (humans who use the process) to select the parts of the process that suit them. There is no evidence of data or validation of such categories in the study.

Nejmeh and Riddle [6] present a Process Evolution Dynamics Framework that allows process change agents to describe, understand, learn from, plan, and manage process evolution efforts. They consider the organization's context as the determinant factor for defining and sequencing process evolution cycles and recommend exploring the context factors that influence process changes in order to better understand process evolution. Customer desires, market pressure, personnel availability, personnel capability, business goals, regulatory constraints, and available technologies are, among others, important business context factors.

Bhuta et al. [7], propose the development of process elements that can be built with reusable strategies, and be reused for creating different project plans. One strategy can be, e.g., to search for a process element, select a process element, understand the process element selected, and, if required, adapt the process element. This means that process elements must be accompanied by important information that can be easily understood by project managers. Examples of such information are: What the process element does, its value, how it could be executed, which resources are required to execute it, and its context information. Butha et al. [7] refer to Basili et al. [8] for the problems of capturing and storing context information in a project repository. Unfortunately, the case study presented by Butha et al. [7] neither provides evidence on context information, nor reasons for selecting certain process elements as part of a project plan.

## Study Context

The study presented in this article was performed in the context of a project that aimed at the evolution of space standards.

The European Cooperation for Space Standardization (ECSS) [13] is an initiative established to develop a coherent, single set of easy-to-use standards for all European space activities, covering all areas of space activities, including engineering, quality assurance, and project management. Organizations or projects part of the European Space Agency (ESA) are supposed to develop and use their specific tailoring(s) of the ECSS standards. Tailoring can be done in a project-specific way (i.e., a separate tailoring for each project) or in an organization-specific way (i.e., one tailoring per organization, to be used for all their projects). The ESA Space Operations Center ESOC (i.e., the ESA organization where the project took place) chose the organization-specific tailoring approach. The applicable implementation of their ECSS tailoring was the Software Engineering and Management Guide (SEMG) [14], which was used for all their major projects.

After some years of experience with the ECSS standards, they were revised by ESA, and a new version was published. This also meant that the SEMG had to be revised, in order to be compliant to the revised ECSS standard. This compliance had to be proven by means of traceability of every ECSS requirement to its implementation, and by providing a tailoring justification for every tailored requirement. The process engineers' task was to tailor the relevant parts of the ECSS (comprising several hundred requirements) to ESOC's needs and to apply this tailoring in an update of their implementation of the standard, the SEMG.

Another important task assigned to process engineers was to improve the ease of use of the SEMG. For the purposes of this project, process engineers considered that the ease of use of a document is positively influenced by improving: (1) internal consistency, i.e., avoiding that one part of the document contradicts another, (2) external consistency, i.e., avoiding that the document at hand contradicts other documents and that links to external sources are correct, and (3) conciseness, i.e., indexed tables of contents allow people to find important things quickly, different concepts are explained and marked clearly, and the document is not larger than necessary.

Finally, process engineers had to maintain detailed change logs on a per-section basis, because of very different stakeholders who wanted to keep track of the changes performed to the SEMG and their justifications.

One initial analysis concerned compliance and showed that the SEMG was only partially compliant to the new ECSS software standard, and had to be updated accordingly. Another initial analysis concerned ease-of-use and was done by analyzing the SEMG documents and by means of structured interviews with SEMG users. Process engineers observed that the most predominant wish was for output simplification and clarification. Furthermore, the SEMG structure did not reflect actual process execution any more and had to be adjusted accordingly.

The SEMG was modified iteratively and incrementally as follows: Process engineers changed the SEMG and delivered a new version for review. Afterwards, reviewers discussed changes performed to the SEMG and accepted or rejected such changes. The reviewers documented their decisions and sent comments and suggestions to the process engineers. Process engineers reworked the SEMG based on the comments and suggestions. This iterative process allowed updating the SEMG in a controlled way and enabled a constant review of the accomplishment of the tasks.

## Data Analysis

Process engineers documented the information related to the changes and their justifications and stored them in a database as they were evolving the SEMG. Two versions of the SEMG resulted from the editing-reviewing iterations. This was an initial attempt at collecting the rationale of process changes in order to understand the nature of changes and to understand how to capture rationale information adequately. The information collected about the changes was used as the basis for a detailed study of the most important and common issues that were resolved by each change. We accomplished this by querying the database that contains information on changes to the SEMG and by understanding each change's justification. While doing this, we derived a list of the most common issues that process engineers faced while doing the SEMG evolution. The following is the list and an explanation of the issues:

1. Improper sequence of processes: Process engineers found that the prescribed control flow of activities differed from the one followed in real projects.
2. Ambiguous activity description: Process engineers found activity descriptions capable of being understood in two or more possible senses or ways.
3. Improper placement of an output: Process engineers found that the prescribed product flow differed from the one present in real projects.
4. Non-compliant activity: Process engineers found cases where activities did not fulfill the requirements stated in the ECSS standards.
5. Ambiguous additional explanatory text: Process engineers found explanatory text that could be understood in two or more possible senses or ways.
6. Improper placement of additional explanatory text: Process engineers found examples of explanations that were incorrectly referenced.

7. Misleading name of an activity: Process engineers found names that did not reflect the meaning of the process for practitioners.
8. Activity description not concise: Process engineers found activity descriptions that contain superfluous or unnecessary statements.
9. Redundant activity description: Process engineers found duplicated descriptions of activities.
10. Additional explanatory text not concise: Process engineers found examples or explanations that contained superfluous or unnecessary statements.
11. Ambiguous output description: Process engineers found output descriptions capable of being understood in two or more possible senses or ways.

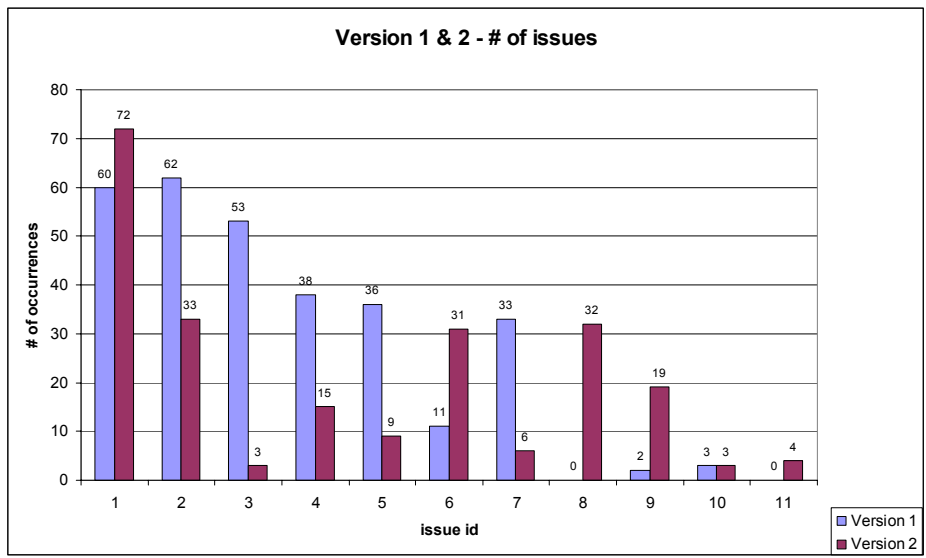

**Figure 1**.  # occurrences per issue

Figure 1 reflects the number of changes caused by the issues listed above when editing the process standards during the first and second iterations. It was found that during the first iteration, issues such as "improper sequence of processes", "ambiguous activity description", and "improper placement of an output", i.e., (1), (2), and (3) respectively, caused the largest number of changes to the process standards. This can be explained by the fact that in the first iteration, the process standards contents and its architecture were extensively modified in order to fulfill the objective of increasing the standard's ease of use.

The next most frequent issue is "non-compliant activity" (4), because one of the goals was to correct the standard's contents so that they were closer to the higher level standard. This leads to the suspicion that process standards were previously evolved without any historical perspective, producing as a consequence standards that totally deviated from the higher level standard.

Figure 1 also reflects the relationship between the issues listed above and the number of changes they caused when editing the process standards during the second iteration. Compared to the first iteration, it can be seen how the number of changes due to "ambiguous activity description" (2) and "non-compliant activity" (4) were reduced more or less to half. Other issues such as "improper placements of an output" (3) and "ambiguous additional explanatory text" (5) were also drastically reduced. This suggests that after the first iteration, process engineers partially accomplished increasing ease of use and compliance to process standards. However, the number of occurrences for issues such as: "improper sequence of processes" (1), "redundant activity description" (9), and "improper placement of additional explanatory text" (6) increased. This can be attributed to the reviewers. Although reviewers were satisfied at the end of the first iteration with the reduced number of "non-compliant activity" issues (4) with respect to the ECSS and less "ambiguous activity descriptions" (2), they still believed that activity descriptions were not correctly grouped. In fact, there were several discussions about the interfaces (inputs and outputs) between system engineering and software engineering processes that demanded a better understanding of the actual practices and reflection in the standards. The reviewers were satisfied concerning the improvement of the process standards at the end of the first iteration and saw the opportunity of having high quality standards at the end of the second iteration. Therefore, they were stricter and demanded higher quality of the process standard contents for the second iteration. This is possibly the reason why new issues appeared such as: "activity description not concise" (8), and "ambiguous output description" (11).

## Summary and Outlook

Processes may be more easily and rationally changed if the information about the process, its context, and the rationale of its evolution is captured. Existing approaches recognize the need for a mechanism (concept and tool) that can be used for collecting information about process changes that could help evolve the process in a systematic way. We observed that most of the approaches did not consider rationale information as an important part of their frameworks. This can be the reason for the small amount of evidence available on the rationale of process evolution. Having a predefined classification of the rationale for process changes, the task of collecting the information related to such rationale can be simplified and become more suitable for use in real process evolution projects. This may be seen as an initial step for building a mechanism that supports systematic process evolution. Therefore, more research effort should be invested into understanding how to introduce these rationale concepts for a systematic well-grounded evolution of software process models. The list of issues derived from analyzing the database with the information about the evolution of process standards provides an initial insight on the type of changes performed in the context of this type of projects. It can be said that the issues that generated the major number of occurrences such as "improper sequence of processes" (1), "ambiguous activity description" (2), and "improper placement of an output" (3), reflected the distance that existed between the process description and the actual understanding of stakeholders. It was observed that systematically documenting changes and discussing them in reviews provided a much more organized and well-grounded process standard evolution. However, a more structured mechanism for collecting the rationale of changes is needed for clearly identifying the observed alternatives and criteria, as well as the arguments and final resolution. More research has to be done for describing more precisely this initial list of issues, so that they are as orthogonal as possible. More empirical data is needed for that purpose. As part of our future work we will use the issues list as the basis for new process evolution projects.

*Acknowledgements.* We would like to thank Michael Jones and Mariella Spada from ESA Space Operations Center (ESOC) and Dr. William E. Riddle for their support and their valuable comments. Additionally, we would like to thank Sonnhild Namingha from Fraunhofer IESE for preparing the English editing of this paper. This work was supported in part by the German Federal Ministry of Education and Research (V-Bench Project, No.01| SE 11 A).

# References

[1]  Dutoit, H, A., Paech, B.: Rationale Management in Software Engineering. Stuttgart: Expected date of publication: Beginning of 2006.
[2]  Nguyen, M, N., Conradi, R.: Towards a rigorous approach for managing process evolution. Software process technology: 5th European workshop, EWSPT '96, Nancy, France. 1996.
[3]  Bruegge, B., Dutoit, A.H.: Object-Oriented Software Engineering. Using UML, Patterns, and Java. 2nd ed. Upper Saddle River: Pearson Education 2004.
[4]  Bandinelli, S., Fugetta, A, Ghezzi, C.: Software Process Model Evolution in the SPADE environment. IEEE Transactions on Software Engineering 19:1128-1144. 1993
[5]  Madhavji, N.: Environment evolution: The Prism model of changes. IEEE Transactions on Software Engineering, 18(5):380-392.
[6]  Nejmeh, Brian A., Riddle, William E.: The PERFECT Approach to Experience-based Process Evolution. Advances in Computers, M. Zelkowitz (Ed.), Academic Press, 2006.
[7]  Bhuta, J., Boehm, B., Meyers, S.: Process Elements: Components of Software Process Architectures. Software Process Workshop, China, (2005).
[8]  Basili V., McGarry F.: The Experience Factory: How to Build and Run One. 19th International Conference on Software Engineering, Boston, Massachusetts, May (1997)
[9]  Kunz, W., Rittel, H.: Issues as Elements of Information Systems. Working Paper No. 131, Institut für Grundlagen der Plannung, Universität Stuttgart, Germany, (1970).
[10] Lee, J.: A Qualitative Decision Management System. In P.H. Winston & S. Shellard (eds.) Artificial Intelligence at MIT: Expanding Frontiers, Vol.1, pp. 104-133, MIT Press, Cambridge, MA, 1990.
[11] MacLean, A., Young, R.M., Belloti, V., Moran, T.: Questions, Options, and Criteria: Elements of Design Space Analysis. Human-Computer Interaction, Vol. 6, pp. 201-250, 1991.
[12] Chung, L., Nixon, B.A., Yu, E., Mylopoulos, J.: Non-Functional Requirements in Software Engineering. Kluver Academic, Boston, 1999.
[13] European Cooperation for Space Standardization (ECSS) Standards available at http://www.ecss.nl. Last checked 2006-01-06.
[14] Ground Segment Tailoring of ECSS for ESOC (SETG), available at http://www.estec.esa.nl/wmwww/EME/Bssc/BSSCdocuments.htm, Last checked 2006-01-06